\documentclass[14pt]{article}
\usepackage[dvips]{graphicx}
\usepackage[dvips]{epsfig}

\date{}

\begin{document}

\begin{center}
 \textbf{ Magnetic Behaviour of Disordered Ising Ferrimagnet in High Magnetic Field}

\vspace{1cm}

\textbf{Sobhendu K.Ghatak} \\
\textrm{ Department of Physics } \\ \textrm{RKMVivekananda University,Belur,Howrah-711202,India}\\

\end{center}

\vspace{0.5cm}

\begin{abstract}

The magnetic behaviour of a disordered ferrimagnetic system $A_pB_{1-p}$ where both $A$ and $B$ represent the magnetic atoms with
respective spin $S_A =1/2$ and $S_B =1$ in presence of high magnetic field is treated theoretically.Assuming the magnetic interaction can
be described through Ising Hamiltonian the approximate free energy is obtained using the cluster-variational method.The field dependence of the  magnetization is then obtained for different concentration $p$ and exchange parameters ($J_{AA}$ , $J_{BB}$ and $J_{AB}$ ).For $p=0.5$,the magnetization $M$ in ferrimagnetic state and in absence of compensation temperature $T_{cm}$ vanishes at $T_C$.Field induced reversal of $M$ is found at switching temperature $T_S$ ($< T_C$) which is decreasing function of field $H$.A maximum in $M$ is found above $T_S$ and the maximum value of $M$ increases with field.In ferrimagnetic state $M$ increases almost linearly at high $H$ region. For system with large ferromagnetic $J_{AA}$,the compensation temperature $T_{cm}$ is increasing function of $J_{BB}$ and $J_{AB}$.The decrease in compensation temperature is linear at small field and tends to saturate at higher field.The sharpness of the magnetization reversal is increased with $H$.For fully compensated state of the system with $p=2/3$,the magnetization in presence of $H$ also exhibits switching behaviour at $T_S$.For  $p=0.2$ the field induced reversal of magnetization occurs more sharply.The orientational switching of the sublattice magnetization $M_A$ and $M_B$ with field increases the Zeeman energy and is the origin of magnetization reversal at $T_S$.

\end {abstract}

\vspace{1.0cm}



\vspace{0.5cm}

\vspace{2.0cm}

\noindent $^*$Corresponding author.

\noindent E-mail address: skghatak@phy.iitkgp.ernet.in

\newpage

\newcommand{\be}{\begin{equation}}
\newcommand{\ee}{\end{equation}}

\section{INTRODUCTION}

Ferrimagnetic state in its simplest form is characterized by an opposing and unequal magnetization of two sublattices below a critical temperature $T_c$. The finite magnetization below $T_c$ results from unequal magnetic moment of constituents metal ions of the material. In addition, differences in rate of thermal demagnetization of sublattice magnetization  can lead to complete cancelation at lower temperature -referred as compensation
temperature $T_{cm}$ that exists in number of ferrimagnetic system [1].The studies of ferrimagnetic materials are normally centered around  their importance in technical applications [2-4].In crystalline lattice the constituents metal ions normally occupy respective sites.On the other hand, the site occupancy tends to be random in disordered lattice. The relative composition of constituent metal ions can also be varied over a wide range in disordered (amorphous) state produced through rapidly quenched method[6,7].The rare-earth -transition metal ferrimagnetic alloys in  amorphous state had been investigated for their potential in magneto-optical recording [8].The real alloy contains, apart from magnetic atom,glass former that stabilizes the disordered state.The amorphous alloy with two kinds of magnetic atom can be, to a first approximation, considered  as binary spin system with two sublattice networks.The field dependence of magnetization in disordered ferrimagnetic materials has been of recent interest,and the magnetization is found to be nearly linear in field at high field region [4,5]. The compensation temperature is expected to be field dependent.\textbf{In this context it is appropriate to examine the field behaviour of the magnetization and the compensation temperature in ferrimagnetic state, and is attempted here based on simple theoretical model.}

Theoretical model frequently utilized to describe the  phase diagram and the magnetic behaviour of disordered magnetic alloy is
disordered  Ising model [9-13].The mixed spin system with two different spins whose interaction is Ising-like is considered as simple model for ferrimagnetic system.Different theoretical methods like the mean-field approximation [14],the effective field theories [15,16], the renormalization-group calculations [17,18] and the Monte-Carlo simulations [19,20] are used to get the phase diagram and critical behaviour of Ising model with spin $S =1/2$ and $S =1$ for two sublattices in ordered lattice. The model has also been studied in different decorated lattices [21,22] and it is predicted that the compensation temperature exists within a specific  region of $J_{AA} - J_{BB}$ plane where $J_{AA} (J_{BB})$
is intra-sublattice exchange interaction in A-(B) sublattice [22].

In this article the results of the field dependence of magnetization in ferrimagnetic state of  mixed  Ising spin system $A_pB_{1-p}$ with $S_A =1/2$ and $S_B =1$ are presented.The approximate procedure as suggested by Oguchi [23] for pure Ising system and extended by Ghatak [13] for disordered Ising system is utilized for evaluation of the free energy.The magnetization,compensation temperature and their field dependence are then obtained from the configuration averaged free energy.In sec.2 the model and the method of calculation are outlined and results are presented in Sec.3.

\section{MODEL AND METHOD OF CALCULATION}

We take a binary alloy $A_pB_q$ of two magnetic atoms A and B
with respective concentration $p$ and $q =1-p$. It is assumed that
all magnetic interactions  are localized and can be described by
Ising Hamiltonian
\begin{equation}
H =  - \sum_{ij} J_{ij}  S_{iz} S_{jz}  -\sum_{i} H_i S_{iz}
\end{equation}

Where $S_{iz}$ is the Ising spin at $i-th$ site and takes the
value $S_A =1/2$ or $S_B = 1$ depending upon the occupation of the site by $A$ or $B$ atom. These values are chosen to reduce
the algebraic complexity .The second term is the Zeeman energy where
the magnetic field $H_i$ (expressed in dimension of energy) at
$i-th$ site is in z-direction. The nearest neighbour exchange
interaction $J_{ij}$ takes value $J_{AA} ,J_{BB}$ and $J_{AB}$ for
the magnetic bond A-A, B-B and A-B respectively. For quenched disordered alloy the free energy $F$, given by [14]
\begin{equation}
F =  -  kT [ \ln Tr \exp ( - \beta H ) ]_{av}
\end{equation}

Where $ [.....]_{av}$ represents the average over all possible alloy
configurations and $\beta =1/kT$. The free energy can be expressed
as
\begin{equation}
F = F_0  - kT  [ \ln < \exp(- \beta V)> ]_{av}
\end{equation}

The quantity $F_0 = -kT  [\ln Tr\exp (-\beta H_0 ) ]_{av}$ refers to the configuration averaged free energy for non-interacting system described by the Hamiltonian $H_0$.The symbol $<...> $ in second term of $F$ represents the ensemble average over the states of $H_0$ and  the operator $V = H - H_0$ .
The non-interacing Hamiltonian $H_0$ is taken as

\begin{equation}
H_0 = \sum_{i}I_i S_{iz}
\end{equation}
Here $I_i$ is parametric field to be determined.\\
For evaluation of the second term of Eq.(3) the approximate procedure used earlier is adopted.The approximation is based on assumption that the system is built out of small independent "
building block".The interaction among spins within the block are treated exactly and the rest of the interaction is represented by  field $I_i$.The field $I_i$ is determined from minimization of the approximate free energy $F_v$ that can be written as [23,13]

\begin{equation}
F_v = F_0 -kT \sum_{l=1}^{L}[\ln<\exp(-\beta V_l)>]_{av}
\end{equation}

Where  $V_l$ is the $l_{th}$  division of $V$ which is divided into
$L$ number of blocks. With decrease of number $(L)$ of division,
$F_v$ tends to exact value of free energy. With the  increase in
size of the block the algebraic complexity grows at faster rate
compared to improvement of the result related to transition
temperature of an Ising model.The building block
consisting of 4-spins for Ising model leads to the results equivalent to  the Bethe approximation. Here the same block is taken for the disordered mixed spin system. The possible atomic configurations for 4-spin block with A or B distributed at lattice points are  shown in Fig. 1. The respective probability of occurrence of the  configuration is noted below  the respective figure.
\begin{figure}[htb]
\begin{center}
\epsfig{file=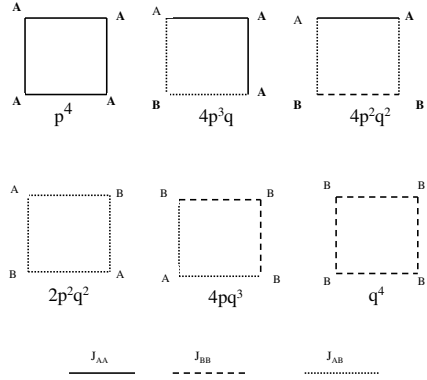,width=10cm}
\caption{Schematic representation of 'building block' of four atoms.
Solid, Dashed   and Dotted lines represent respectively
$J_{AA}$,$J_{AB}$ and $J_{BB}$. The probabilities of different
configurations are given below  the  diagrams.} \label{fig:1}
\end{center}
\end{figure}
The number of spin states  in a given configuration depends on
number of A and B atoms. The number varies from maximum $3^4$  for
block with all $S =1$ atoms to  minimum $2^4$ for all A-atom block.
In this approximation the configuration - averaged trial free energy
$F_v$ can be expressed as [24]
\begin{equation}
F_v =  F_0 - (z/8\beta) [ p^{4}\ln Z_0  + q^{4}\ln Z_4  +4 p^{3}q\ln
Z_1 + 4pq^{3}\ln Z_3 + 2 p^{2}q^{2}2 (\ln Z_2 + 2\ln Z_{22}) ]
\end{equation}
where
\begin{equation}
F_0=-((1-z/2)/\beta)[ p \ln (\cosh (\beta I_A/2))+ q \ln
     (1+2\cosh (\beta I_B)) ]
\end{equation}
and
\begin{equation}
Z_0= 2[ X_A^{4}\cosh 2\alpha_1 +4 \cosh \alpha_1 + 2+ X_A^{-4}]
\end{equation}
\begin{eqnarray}
Z_4&=& 2 X_B^{4} \cosh 4\alpha_2 +8 X_B^{2} \cosh 3\alpha_2
      + 4 (3 +2 X_B) \cosh 2\alpha_2\nonumber\\
   &+& 8 (3+X_B^{2})\cosh \alpha_2
      + 8 X_{BB}^{-1}+ 9 +2X_B^{-4}
\end{eqnarray}
\begin{eqnarray}
Z_1 &=& 2  X_A^{2}X_C \cosh(\alpha_2 +3\alpha_1 /2)
     +2X_A^{2}X_C^{-1}\cosh(-\alpha_2 +3\alpha_1 /2)
    + 2 [2+X_A^{-2}X_C ]\cosh(\alpha_2 +\alpha_1/2)\nonumber\\
      &+&2[2+X_A^{-2}X_C^{-1}]\cosh(-\alpha_2 +\alpha_1/2)
     + 2 [2+X_A^{-2}]\cosh(\alpha_1/2)+ 2 X_A^{2}\cosh(3\alpha_1
     /2)
\end{eqnarray}
\begin{eqnarray}
Z_3 &=& 2 X_B^{2}X_C \cosh(3\alpha_2+\alpha_1/2)
    +2 X_B^{2}X_C^{-1}\cosh( 3\alpha_2 -\alpha_1/2)\nonumber\\
    &+& 2[3+X_B^{-2}X_C+2X_C^{1/2}]\cosh(\alpha_2+\alpha_1/2)
    +2[3+X_B^{-2}X_C^{-1}+2X_C^{1/2}]\cosh(\alpha_2-\alpha_1/2)\nonumber\\
    &+&2[2X_B X_C^{1/2} + X_C ] \cosh(2\alpha_2+\alpha_1/2)
    +2[2X_BX_C^{-1/2} + X_C^{-1}]\
    \cosh(2\alpha_2
    -\alpha_1/2)\nonumber\\
    &+&[6+4(X_C+X_{-C})X_B^{-1}]\cosh(2\alpha_1/2)
\end{eqnarray}
\begin{eqnarray}
Z_2 &=& 2 X_C^2\cosh(2\alpha_2 +\alpha_1) +2
      X_C^{-2}\cosh(2\alpha_2-\alpha_1)+4\cosh(2\alpha_2)+6\cosh(\alpha_1)\nonumber\\
    &+&8\cosh(\alpha_2)+6+4 X_C \cosh(\alpha_2 +\alpha_1) +4
     X_C^{-1}\cosh(\alpha_2-\alpha_1)
\end{eqnarray}
\begin{eqnarray}
Z_{22} &=& 2 X_AX_BX_C \cosh(2\alpha_2 +\alpha_1) +2 X_AX_BX_C
         ^{-1}\cosh(2\alpha_2 - \alpha_1)
         +4X_BX_A^{-1}\cosh(2\alpha_2)\nonumber\\
         &+&2X_A [1+2X_B^{-1}] \cosh(\alpha_1)+
         4X_A X_C^{1/2}\cosh(\alpha_2 +\alpha_1)+4 X_AX_C^{-1/2}\cosh(\alpha_2-\alpha_1)
         )\nonumber\\
         &+&2[2X_A^{-1}(X_C^{-1/2}+ X_C^{1/2}]\cosh(\alpha_2)
         +2X_A^{-1}+2X_A^{-1}X_B^{-1}(X_C^{-1} + X_C)
\end{eqnarray}
\begin{equation}
\alpha_1 = [I_A (1-2/z) +2 H_A/z]\beta,\alpha_2=[I_B(1-2/z)+
2 H_B/z ]\beta
\end{equation}
where z =number of nearest neighbours, $ X_A = \exp (\beta
J_{AA}/4)$ , $X_B =\exp (\beta J_{BB})$ and $ X_C = \exp(\beta
J_{AB})$.
In above expressions of $\alpha$ 's the symbols  $H$'s and $I$'s
represent respectively the magnetic field and the variational
parameters at respective site.\\
The equations for the variational parameters $I_A$
and $I_B$ are obtained from the minimization of the
trial free energy $F_v$
 \begin{equation}
dF_v/dI_A  =  dF_v/dI_B  = 0
\end{equation}
This leads to  the coupled  equations  for $I_A$  and $I_B$  as
\begin{equation}
2\tanh(\beta I_A/2) =  p^3(A_0/Z_0) + 4p^2q(A_1/Z_1) + 4q^3(A_3/Z_3)
+2 pq^2{(A_2/Z_2) + 2(A_{22}/Z_{22})}
\end{equation}
\begin{equation}
\frac{8\sinh(\beta I_B)}{1+ 2\cosh(\beta I_B)} = q^3(B_4/Z_4)+
4p^3(B_1/Z_1)+4pq^2(B_3/Z_3)+2p^2q((B_2/Z_2)+2(B_{22}/Z_{22)})
\end{equation}
where $A_i= (2\beta/z)\frac{\partial Z_i}{\partial H_A}$  and $B_i= (2\beta/z)\frac{\partial Z_i}{\partial H_B}$.The non-trivial self-consistent solutions of the equs. (16-17 )then provide
the approximate free energy.The sub-network magnetization per atom then becomes
\begin{equation}
M_A  =  0.5\tanh (\beta I_{A0} / 2)
\end{equation}
\begin{equation}
M_B  =  2\sinh(\beta I_{B0}) /[1 + 2 \cosh(\beta I_{B0})]
\end{equation}
\\and the total magnetization $M$ per atom
\begin{equation}
 M = p M_A + q M_B .
\end{equation}
Here $I_{A0}$  and $I_{B0}$ are  the  self-consistent solution of
the coupled equations (16) and (17). The finite values of $I_{A0}$
and $I_{B0}$ lead to spontaneous sub-network magnetization which
appears below the transition temperature $T_c$ [24].The numerical results for magnetization are presented for $p =0.5,2/3$ and $0.2$ by varying magnetic field.The model parameters e.g  $kT$, magnetic field $H$ and the exchange interactions are scaled in terms of strongest exchange parameter and number of nearest neighbour is taken as $z=8$.The direction of the magnetic field is assumed to be parallel to the magnetization of $A$-sub-lattice.

\section{RESULTS AND DISCUSSIONS}

\subsection{Magnetization}
\textbf{(i)\ Concentration $p =0.5$} \\
We first examine the ferrimagnetic behaviour of the system with equal concentration of $A$ and $B$ and the exchange interaction between $A$ and $B$ is anti-ferromagnetic.It is also assumed that the exchange integral $J_{AB}$ is stronger than the ferromagnetic exchange between $A-A$ and $B-B$. To represent this case, a typical values  $J_{AB} = -1.0$, $J_{AA} = 0.2 $ and $J_{BB} = 0.1$ for the exchange interactions are taken.The spontaneous
magnetization $M_A $,$M_B$ of $A$ - and $B$ - sublattice and the net magnetization $M$ decrease smoothly with $T$ from their maximum value at $T=0$ and vanish at critical temperature $T_c$ (Fig.-2). At low $T$,
\begin{figure}[htb]
\begin{center}
    \epsfig{file=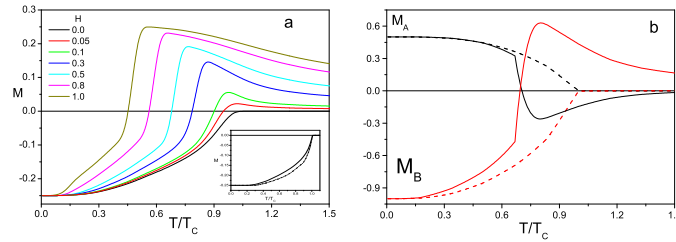,width=10cm}
    \caption{a) Variation of net magnetization $M$ for $p=0.5$ with reduced temperature $T / T_C$ for different field $H$. b) that of sublattice magnetization $M_A$ and $M_B$ for $H = 0$ (dotted line) and $0.5$ (solid line). Value of exchange parameters are $J_{AB} = -1.0,J_{AA} = 0.2$ and $ J_{BB} = 0.1$.Inset Fig.: $M/M_0$ vs $T/T_C$ for ordered $AB$ (dashed curve) and disordered (solid curve) case with same exchange parameters }
\end{center}
\end{figure}
$M$ varies little from its maximum value $M_0= p M_A+(1-p) M_B = -0.25$. \textbf{There is no compensation point below $T_C$ for this situation.In presence of a magnetic field along z-direction, the thermal demagnetization becomes faster for $T \leq T_c$ and changes its sign at a particular temperature $T= T_S(< T_c)$(Fig.$2a$) which depends on the magnetic field.Above $T> T_S$,the magnetization continues to increase until it reaches a maximum at a temperature slightly higher than $T_S$.With further increase in $T$ a slower decrease in the magnetization is found.As the field increases the reversal occurs at lower temperature ,and so $T_S$, termed as the switching temperature, decreases with the field.The maximum value of the magnetization ($\Delta M$) also becomes higher.This large change in behaviour of $M$ near $T_S$ is the result of a different thermal behaviour  of the sublattice magnetization in presence of a moderate field.The results of $M_A$ and $M_B$ are presented in Fig.$2b$ for $H=0$ (dotted) and $0.5$ (solid) and show that a simultaneous alteration of the orientation of sublattice magnetization with respect to the field direction occurs at $T_S$.The switching of the orientation of the magnetization is the result of combined effects of the Zeeman and the exchange energies.}As the magnetic moment of $B$ is higher than that of $A$, the free energy in the switched state is lowered by increasing the magnitude of the Zeeman energy without any loss in the exchange energy.We also note that apart from the field,the inter-sublattice exchange parameters alter the switching temperature,the sharpness of switching and the maximum value of $M$.\textbf{Inset Fig.2a shows thermal variation of the reduced magnetization (dashed curve)of an ordered ferrimagnet where two interpenetrating sublattices $A$ and $B$ are antiferromagnetically aligned.The corresponding variation of $ M/M_0$ for disordered case is given by solid curve.The magnetization at intermediate temperature interval in disordered system is reduced compared to that of ordered state.Close to transition temperature the magnetization are nearly equal in both cases.This is expected when the correlation length around $T_C$ is undisturbed by disorder.Similarity of the magnetization at low temperature is related to the absence of the spin wave excitation.The reversal of the magnetization in ordered system is found to occur at higher field.}
\begin{figure}[htb]
\begin{center}
    \epsfig{file=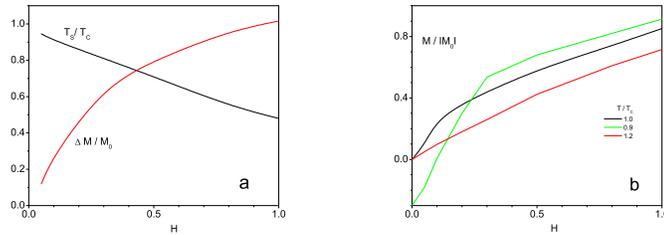,width=10cm}
    \caption{a) Variation of $\Delta M /|M _0|$ and $T_S / T _C$ with $H$ and b) $H$ dependence of magnetization $M /M_0$ at $ T /T_C = 1,0.9$ and $1.2$ for same set of $J$'s as in Fig.2 and $p=0.5$.}
\end{center}
\end{figure}

In Fig.$3a$  the field variation of the reduced switching temperature $T_S/T_c$ and normalized $\Delta M/|M_0|$ are depicted for range $H = 0$ to $1$. In this range $\Delta M/|M_0|$ sharply increases at lower range of $H$ and tends to saturate at higher region.On the other hand, the reduced switching temperature $T_S /T_c$ exhibits nearly linear decrease at high $H$ region.The field dependence of reduced $M/|M_0|$  is also shown in Fig.$3b$ for $T / T_c =1.2 ,1.0$ and  $0.9$.At $T_c$ , $M$ grows faster at low field ,and the growth rate slows down at higher field. For $T < T_c $, $M$ changes sharply at certain field that depends on $T$. In high field region, a nearly linear dependence of $M$ on $H$ is found,and the slope becomes smaller at higher $T$ due to more thermal fluctuation.\\

Next we consider a system where the exchange parameters are such that the $J_{AA} $ is largest compared to others. This is commonly the situation for the rare-earth – transition metal alloy where the exchange integral between the transition metal ions is much stronger than that between rare-earth and transition metal ions or between rare-earth ions.The moment of the rare-earth ($B$) is larger than that of transition metal ion ($A$).Again, there are situations where $J_{AB} $ is not very weak compared to $J_{AA} $.In Fig.$4$ the magnetization $M$ ($4a$) and the sublattice magnetization $M_A$ and $M_B$ ($4b$) are displayed for the exchange parameters $J_{AA} =1.0,J_{AB} = -0.5$, and $J_{BB} = 0.05$.
\begin{figure}[htb]
\begin{center}
    \epsfig{file=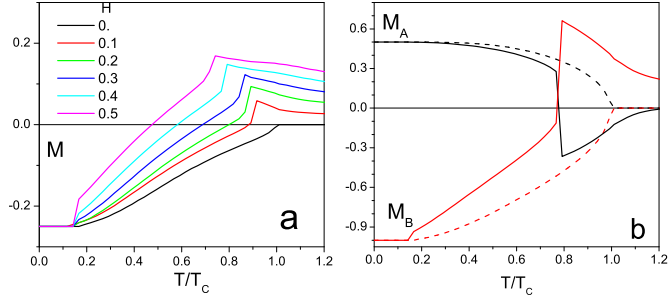,width=10cm}
    \caption{a) Variation of net magnetization $M$ for $p=0.5$ with reduced temperature $T / T_C$ for different field $H$.b) that of sublattice magnetization $M_A$ and $M_B$ $H = 0$ (dotted line) and $0.4$ (solid line). Value of exchange parameters are $J_{AA} = 1.0,J_{AB} = -0.5,$ and $ J_{BB} = 0.05$ }
\end{center}
\end{figure}
The net magnetization  $M$ in the ferrimagnetic phase at $H=0$ varies little for small $T / T_C$.However,a larger variation is found at higher temperature due to faster demagnetization of $M_B$.This is associated with smaller inter- and intra-exchange $J_{BB}$ interactions.
For this set of parameters there is no compensation temperature and $M$ vanishes at $T_C$.In presence of a magnetic field $M$ changes
sign at a temperature $T=T_S$ (field –induced switching temperature) and attains a maximum at a temperature $T_M $ less than $T_C$. The reversal of $M$ at $T_S$ is the effect of higher Zeeman energy gain when $M_A$ and $M_B$ switches their orientation with respect to the magnetic field (Fig.4b). The field variation of the maximum value of the magnetization with respect to its maximum magnitude at zero field $M_M/ |M_0|,$, is given in Fig.$5$.A sharp increase is
\begin{figure}[htb]
\begin{center}
\epsfig{file=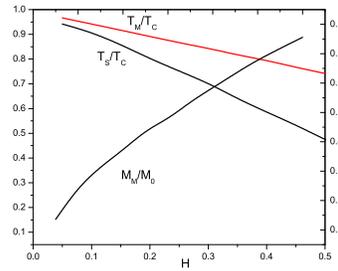,width=5cm}
\caption{ Field variation of  the maximum value of $M_M/|M_0|$,the normalized switching  temperature $T_S/T_C $ and the temperature $T_M /T_C$ where $M$ is maximum.Other parameters are same as in Fig.4. }
\end{center}
\end{figure}
found at low $H$ and tends to saturate at higher field region.With increase in $H$ both  $T_M$ and $T_S$ (normalized by $T_C$) are reduced.The variation is almost linear for $T_M$ whereas nonlinearity is evident for $T_S$.When the inter-sublattice interaction is further reduced then the compensation temperature appears before the transition temperature $T_C$.One such situation is described by the Fig.$6$ for a set of exchange $J_{AA} = 1.0,J_{AB} = -0.1,$ and $ J_{BB} = 0.05$ for different
\begin{figure}[htb]
\begin{center}
    \epsfig{file=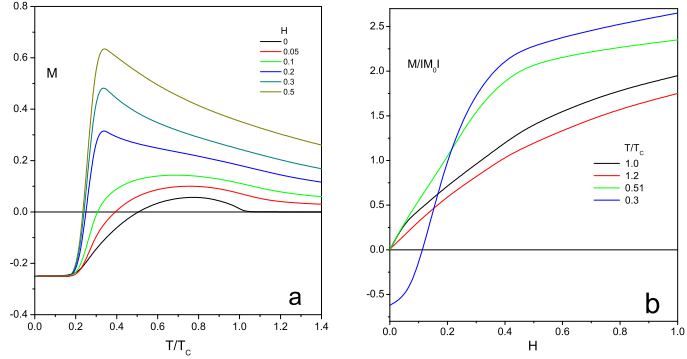,width=10cm}
    \caption{ Variation of net magnetization $M$ for $p=0.5$ with reduced temperature $T / T_C$ for different field $H$ (a),and the reduced magnetization $M/|M_0|$ with field $H$ (b) for different values of $T / T_C$.Values of exchange parameters are $J_{AA} = 1,0,J_{AB} = -0.1,$ and $ J_{BB} = 0.05$ }
\end{center}
\end{figure}
field ranging from $0$ to $0.5$.At low temperature  $M$ is again dominated by $B$ sublattice magnetization, changes its sign at the compensation temperature $T_{CM}$ and finally vanishes at $T_C$. At compensation point the magnetization of whole system becomes zero due to equal and opposite magnetization of two sub-lattices. A faster demagnetization of sublattice $B$  compared to $A$ sublattice occur as both  $J_{AB} = -0.1,$ and $ J_{BB} = 0.05$ are much weaker compared to $J_{AA}$.In presence of $H$ the compensation occurs at lower temperature and a well defined maximum appears (Fig.$6a$).The temperature dependence is also highly asymmetric about temperature where the maximum appears. For high field the reversal of $M$ is sharp around $T{CM}$.\textbf{In Fig.$6b$ the field dependence of  $M$ is shown for temperature $T/T_C =1.2,1,0.51$ and $0.3$.The curve labeled $0.51$ corresponds to the result at compensation point $T_{CM} =0.51T_C$ when $H=0$.The curve labeled $0.3$ is the result when temperature is selected as $T<T_{CM}$.The magnetization at low temperature $T<T_{CM}$ changes its direction at faster rate.At $T_{CM}$,$M$ grows with high slope for small field,however the variation of $M$ becomes flattened at  a higher field. Nearly linear variation is observed at high field although the rate of variation depends on temperature. The compensation temperature $T_{CM}$ at $H=0$ depends on exchange parameters $J_{BB}$ and $J_{AB}$ [24]. The compensation point in ferrimagnetic state appears when the sublattice $B$ with higher moment is thermally demagnetized at a faster rate compared to that of other sublattice $A$.This happens when intra-sublattice interaction $J_{BB}$ is weak.For given $J_{AB}$,it is expected that higher value of ferromagnetic $J_{BB}$ results higher $T_{CM}$.The dependence of $T{CM}$  on $J_{BB}$ for two values of $J_{AB} = -0.1$}
\begin{figure}[htb]
\begin{center}
    \epsfig{file=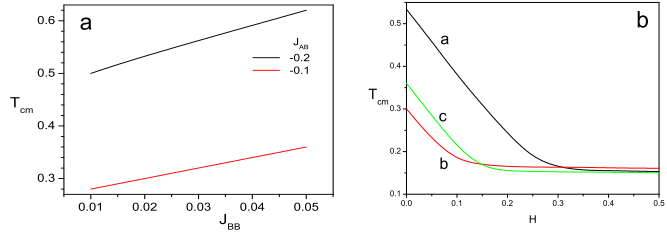,width=10cm}
    \caption{a)Dependence of the compensation temperature $T_{CM}$ with exchange $J_{BB}$ for two values of $J_{AB}$ and $=0$.b) Field dependence of $T_{CM}$ for three set of exchange parameters a) $J_{AB} = -0.2,J_{BB} =0.02$, b)$J_{AB} = -0.1 , J_{BB} =0.02$,c)$J_{AB} = -0.1 , J_{BB} =0.05$  and $p=0.5$}
\end{center}
\end{figure}
and $-0.2$ is displayed in Fig.$7a$ ,and  $T{CM}$  is linearly dependent on $J_{BB}$.For $J_{AB} =-0.2$ the compensation point exists when  $J_{BB} $is less than $0.1$.The dependence of $T{CM}$ on $H$ is given in Fig.$7b$ for $J_{AA} =1.0$ and a) $J_{AB} = - 0.2$ ,$J_{BB} =0.02$, b) $J_{AB} = - 0.1$,$J_{BB} =0.02$,and  c) $J_{AB} = - 0.1$,$J_{BB} =0.05$,$J_{AB} = 0.1$. For low $H$, a linear decrease in $T{CM}$ is found for all cases,and at higher field $T{CM}$ tends to saturate.

\textbf{(i)\ Concentration $p = 2/3$} \\
The concentration $p=2/3$ corresponds to the situation where total magnetization is completely compensated at $T=0$ due to choice of spin values. When the inter-sublattice exchange $J_{AB} = -1.0$ is larger than the intra-sublattice exchange  ($J_{AA} =  J_{BB} = 0.4$)interactions  the state with $M = 0$ persists up to $T/TC =0.2$ (Fig.$8a$).Within intermediate temperature interval ( $ 0.2\leq T < T_C$) and $H=0$, $M$ becomes finite  but small and dominated by $B$-sublattice magnetization as thermal demagnetization effect is same for both sublattices.However,the presence of field changes the behavior of $M$. There is a sharp variation of $M$ from negative to positive value at switching temperature as $H$ increases. With increases in field the switching temperature $T_S$ is lowered.The  switching like behaviour of $M$ is due to simultaneous flipping of sublattice magnetization in order to maximize Zeeman energy and keeping antiferromagnetic alignment of the sublattice magnetizations.The field dependence at high field is also nearly linear as in earlier case of $p = 0.5$. We note that the magnitude of change of the magnetization in this case much smaller compared to the system $p = 0.5$.
\begin{figure}[htb]
\begin{center}
    \epsfig{file=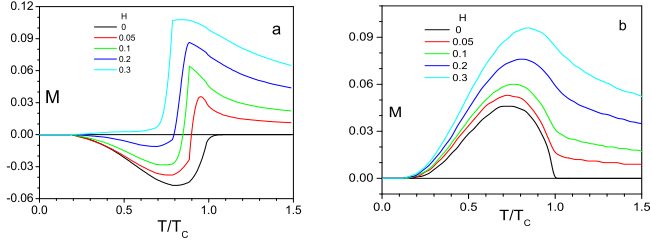,width=10cm}
    \caption{ Variation of net magnetization $M$ for $p=2/3$ with reduced temperature $T / T_C$ for different field $H$. Value of exchange parameters are for a)$J_{AA} = 0.4,J_{AB} = -1.0,$ and $ J_{BB} = 0.4$  and for b) $J_{AA} = 1.0,J_{AB} = -0.5,$ and $ J_{BB} = 0.2$ }
\end{center}
\end{figure}
On the other hand when  $J_{AA} = 1.0$ dominates over other exchange interactions ($J_{AB} = -0.5 ,J_{BB} -0.2$) (Fig.$8b$) the magnetization always remain parallel to $H$.This is due larger exchange energy in sublattice $A$ compared to that in $B$ sublattice, and the induce magnetization is dominated by change in $M_A$ for all field.

\textbf{(i)\ Concentration $p = 0.2$} \\
This refers to the situation where the concentration of ions with higher values of magnetic moment is smaller and can simulate amorphous transition metal -rare-earth alloy with smaller concentration of rare-earth.Fig.$9a$ displays the net magnetization $M$  for different field $H= 0$ to $0.5$ for exchange  $J_{AA} = 1.0,J_{AB} = -0.5,$ and $ J_{BB} = 0.2$.There is no compensation temperature for this set of parameters.But the reversal occurs in presence of field.It is found that $M$ switched very sharply from negative to positive value.Again the switching behaviour is associated with reversal of direction of $M_A$ and $M_B$ with respect to the field direction at $T_S$.
\begin{figure}[htb]
\begin{center}
    \epsfig{file=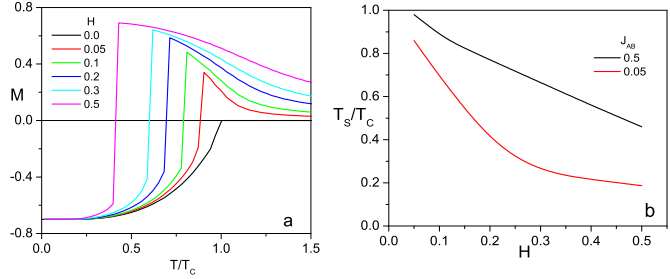,width=10cm}
    \caption{ Variation of net magnetization $M$ for $p=0.2$ with reduced temperature $T / T_C$ for different field $H$(a).Value of the exchange parameters are $J_{AA} = 1.0,J_{AB} = -0.5,$ and $ J_{BB} = 0.2$.b)Field dependence of $T_S/T_C$ for two values of $J_{AB}$.}
\end{center}
\end{figure}

\textbf{ In this work it is assumed that the magnetization of the sub-lattice $A$ carrying smaller moment per site is aligned parallel to +z-axis and antiferromagnetically to the magnetization of sub-lattice $B$ with higher moment.This leads to net magnetization is in negative z-direction.In real system the direction of the magnetization is determined by an anisotropy energy which often is represented by anisotropic field acting on sub-lattice. So by adding small anisotropic field along +z direction in sub-lattice $A$ the presented ferrimagnetic state is realized.The switching behaviour of the magnetization in the ferrimagnetic state with magnetic field is expected to follow when the field is in opposite direction of the magnetization.The switching of magnetization had been observed in multilayer of rare-earth and transition metal[25,26] or alloy[26].The magnetization of rare-earth layers is aligned opposite to that of transition metal layers due to antiferromagnetic interaction at the interface of two layers.The magnetization reverses its direction in presence of the magnetic field and exhibits complex hystersis.Although these systems are different compared to system considered here however, it is worth noting the similarity of global field behaviour of the magnetization.We envisage possible application of the field induced magnetization reversal.The switching  device that relied on the sign of the magnetization is a possible area of application.At a fixed field the switching will be induced by changing temperature through $T_S$.The ferrimagnetic system with higher concentration of ion carrying high magnetic moment would be more appropriate due to sharp nature of reversal (Fig.9).}

\section{CONCLUSIONS}

A disordered ferrimagnetic alloy ($A_pB_{1-p}$) with Ising like interaction between the spins ($S_A=1/2 $and $ S_B=1$) of
sub-lattices is treated using a cluster-variational method in presence of magnetic field.In this method the interactions within the cluster of different configurations are treated exactly and the rest of the interaction is described by variational field which is obtained from
minimization of free energy. The results on the magnetization and compensation temperature are presented for $p = 0.5,2/3$ and $0.2$ for different values of field and exchange parameters.In absence of a compensation temperature at zero field,a field induced magnetization reversal in ferrimagnetic state is found at a temperature -termed as switching temperature.With increase in the magnetic field the switching occurs at lower temperature and the magnetization reversal becomes sharper.At switching temperature the free energy is gained by interchanging orientation of sublattice magnetization with respect to applied field.The compensation temperature at zero field increases with increase in intra-sublattice exchange interaction of $B$ sublattice.In presence of magnetic field the the compensation point appears at smaller temperature.The magnetization in paramagnetic state  varies almost in linear fashion with $H$ in high field region.The nearly linear increase in magnetization has been observed in rare-earth - transition metal alloy [4,5].For fully compensated composition ($p = 2/3$) the magnetization reverses at a switching temperature when inter-sublattice exchange dominates over others.On contrary the magnetization passes through a maximum  at $T < T_C$ when the intra-sublattice exchange is largest.The magnetization reversal is found to be much sharper for system with higher concentration of $B$ with weak $J_{AB}$.The effect of field induced magnetization reversal can be utilized for switching device.

\section{Acknowledgement} The author gratefully acknowledges assistance from authority of RKMVivekananda University.It is also my great pleasure to acknowledge Prof.D.Sherrington for his help and encouragement during work on disordered spin system.

\end{document}